\documentclass[aps,preprint]{revtex4-1}

\usepackage{bm}
\usepackage{graphicx}
\usepackage{color}
\usepackage[normalem]{ulem}
\usepackage{epstopdf}
\usepackage{amsmath}
\usepackage{amssymb}
\usepackage{soul}
\usepackage[capitalize]{cleveref}
\usepackage{color}
\usepackage{enumitem}

\begin{document}

\title{Imaging vibrations of locally gated, electromechanical few layer graphene resonators with a moving vacuum enclosure}

\author{Heng Lu}
\affiliation{School of Optoelectronic Science and Engineering \& Collaborative Innovation Center of Suzhou Nano Science and Technology, Soochow University, Suzhou 215006, People's Republic of China}
\affiliation{Key Lab of Advanced Optical Manufacturing Technologies of Jiangsu Province \& Key Lab of Modern Optical Technologies of Education Ministry of China, Soochow University, Suzhou 215006, People's Republic of China}

\author{Chen Yang}
\affiliation{School of Optoelectronic Science and Engineering \& Collaborative Innovation Center of Suzhou Nano Science and Technology, Soochow University, Suzhou 215006, People's Republic of China}
\affiliation{Key Lab of Advanced Optical Manufacturing Technologies of Jiangsu Province \& Key Lab of Modern Optical Technologies of Education Ministry of China, Soochow University, Suzhou 215006, People's Republic of China}

\author{Ye Tian}
\affiliation{School of Optoelectronic Science and Engineering \& Collaborative Innovation Center of Suzhou Nano Science and Technology, Soochow University, Suzhou 215006, People's Republic of China}
\affiliation{Key Lab of Advanced Optical Manufacturing Technologies of Jiangsu Province \& Key Lab of Modern Optical Technologies of Education Ministry of China, Soochow University, Suzhou 215006, People's Republic of China}

\author{Jun Lu}
\affiliation{School of Optoelectronic Science and Engineering \& Collaborative Innovation Center of Suzhou Nano Science and Technology, Soochow University, Suzhou 215006, People's Republic of China}
\affiliation{Key Lab of Advanced Optical Manufacturing Technologies of Jiangsu Province \& Key Lab of Modern Optical Technologies of Education Ministry of China, Soochow University, Suzhou 215006, People's Republic of China}

\author{Fanqi Xu}
\affiliation{School of Optoelectronic Science and Engineering \& Collaborative Innovation Center of Suzhou Nano Science and Technology, Soochow University, Suzhou 215006, People's Republic of China}
\affiliation{Key Lab of Advanced Optical Manufacturing Technologies of Jiangsu Province \& Key Lab of Modern Optical Technologies of Education Ministry of China, Soochow University, Suzhou 215006, People's Republic of China}

\author{FengNan Chen}
\affiliation{School of Optoelectronic Science and Engineering \& Collaborative Innovation Center of Suzhou Nano Science and Technology, Soochow University, Suzhou 215006, People's Republic of China}
\affiliation{Key Lab of Advanced Optical Manufacturing Technologies of Jiangsu Province \& Key Lab of Modern Optical Technologies of Education Ministry of China, Soochow University, Suzhou 215006, People's Republic of China}

\author{Yan Ying}
\affiliation{School of Optoelectronic Science and Engineering \& Collaborative Innovation Center of Suzhou Nano Science and Technology, Soochow University, Suzhou 215006, People's Republic of China}
\affiliation{Key Lab of Advanced Optical Manufacturing Technologies of Jiangsu Province \& Key Lab of Modern Optical Technologies of Education Ministry of China, Soochow University, Suzhou 215006, People's Republic of China}

\author{Kevin G. Sch\"adler}
\affiliation{ICFO--Institut de Ciencies Fotoniques, The Barcelona Institute of Science and Technology, 08860 Castelldefels Barcelona, Spain}

\author{Chinhua Wang}
\affiliation{School of Optoelectronic Science and Engineering \& Collaborative Innovation Center of Suzhou Nano Science and Technology, Soochow University, Suzhou 215006, People's Republic of China}
\affiliation{Key Lab of Advanced Optical Manufacturing Technologies of Jiangsu Province \& Key Lab of Modern Optical Technologies of Education Ministry of China, Soochow University, Suzhou 215006, People's Republic of China}

\author{Frank H. L. Koppens}
\affiliation{ICFO--Institut de Ciencies Fotoniques, The Barcelona Institute of Science and Technology, 08860 Castelldefels Barcelona, Spain}

\author{Antoine Reserbat-Plantey}
\email{Antoine.Reserbat-Plantey@icfo.eu}
\affiliation{ICFO--Institut de Ciencies Fotoniques, The Barcelona Institute of Science and Technology, 08860 Castelldefels Barcelona, Spain}

\author{Joel Moser}
\email{j.moser@suda.edu.cn}
\affiliation{School of Optoelectronic Science and Engineering \& Collaborative Innovation Center of Suzhou Nano Science and Technology, Soochow University, Suzhou 215006, People's Republic of China}
\affiliation{Key Lab of Advanced Optical Manufacturing Technologies of Jiangsu Province \& Key Lab of Modern Optical Technologies of Education Ministry of China, Soochow University, Suzhou 215006, People's Republic of China}

\begin{abstract}
Imaging the vibrations of nanomechanical resonators means measuring their flexural mode shapes from the dependence of their frequency response on in-plane position. Applied to two-dimensional resonators, this technique provides a wealth of information on the mechanical properties of atomically-thin membranes. We present a simple and robust system to image the vibrations of few layer graphene (FLG) resonators at room temperature and in vacuum with an in-plane displacement precision of $\approx0.20$~$\mu$m. It consists of a sturdy vacuum enclosure mounted on a three-axis micropositioning stage and designed for free space optical measurements of vibrations. The system is equipped with ultra-flexible radio frequency waveguides to electrically actuate resonators. With it we characterize the lowest frequency mode of a FLG resonator by measuring its frequency response as a function of position on the membrane. The resonator is suspended over a nanofabricated local gate electrode acting both as a mirror and as a capacitor plate to actuate vibrations at radio frequencies. From these measurements, we estimate the ratio of thermal expansion coefficient to thermal conductivity of the membrane, and we measure the effective mass of the lowest frequency mode. We complement our study with a globally gated resonator and image its first three vibration modes. There, we find that folds in the membrane locally suppress vibrations.
\end{abstract}

\maketitle

\section{Introduction}

Imaging the flexural vibrations of two-dimensional (2-D) nanomechanical resonators is an important task. These resonators, made of atomically-thin membranes of graphene \cite{graphene1,Changyao,Craighead,graphene2,graphene3,Mandar,GuoPing} and various thin materials such as few layer transition metal dichalcogenides \cite{NbSe2,TMD1,embracing,TMD14,Tantalum}, offer the opportunity to study the physics of vibrational modes in regimes where extremely small mass, low bending rigidity, large stretching rigidity and large aspect ratio combine to give rise to a wealth of mechanical behaviors \cite{Mark}. In its simplest form, imaging vibrations means measuring their time averaged, resonant amplitude as a function of position on the membrane. Driven vibrations of graphene resonators were imaged with an Atomic Force Microscope (AFM) \cite{Daniel} and with an optical interferometry setup \cite{Craighead}. Both driven and thermal vibrations of resonators based on graphene \cite{Davidovikj,Antoine}, MoS$_2$ \cite{embracing}, black phosphorus \cite{blackP}, and hexagonal boron nitride \cite{HBN} were measured using a similar optical interferometry technique. These measurements advanced our understanding of 2-D resonators in several important ways. They made it possible to identify modes in the vibration spectrum unambiguously, including degenerate modes that are otherwise difficult to detect \cite{Craighead,Davidovikj}. They revealed the impact of unevenly distributed stress and mass on the mode shape \cite{Daniel}. They also proved to be exquisitely sensitive to mechanical anisotropies, such as the anisotropy of Young's modulus that stems from the crystal structure of the membrane \cite{blackP}. All these interesting measurements may benefit nanomechanical sensing applications, including spatially resolved nanomechanical mass spectroscopy \cite{Roukes}. Vibration imaging may also complement other techniques including AFM, Raman spectroscopy and photoluminescence where these are used to inspect 2-D materials for defects, impurities and grain boundaries.

With all its merits, imaging vibrations of 2-D resonators remains a challenging task. Difficulties come in part from the necessity to measure vibrations in a controlled environment. Measuring the mechanical response of thin membranes while minimizing their damping rates requires keeping them in vacuum. This immediately places technical constraints on the measuring equipment. Piezo linear actuators \cite{Precision1,Precision2} were used in some imaging experiments, where they either moved the vacuum enclosure \cite{Davidovikj} or moved a microscope objective with respect to the vacuum enclosure \cite{Antoine}, while a high-precision motorized stage was used in other experiments \cite{embracing,blackP,HBN}. Adding to the complexity of imaging vibrations, radio frequency electrical signals are sometimes supplied to the resonator to drive vibrations while the position of the resonator is changing. Implementing this driving technique is difficult because radio frequency cables are stiff and hinder the motion of piezoelectric actuators and motorized stages.

Here we demonstrate vibration imaging of electrically driven, few layer graphene (FLG) resonators using a moving vacuum enclosure. The enclosure is mounted on a three-axis micropositioning stage with a measured in-plane displacement precision of $\approx0.20$~$\mu$m and is equipped with homemade, ultra-flexible radio frequency waveguides. Our design makes use of the large load capacity of the stage and the high sensitivity of the manual adjusters attached to it. The adjusters are driven by inexpensive stepper motors connected to them by simple gears and rubber belts. The benefits of our system are its sturdy design that protects it against acoustic vibrations, its submicrometer displacement precision unaffected by a rather heavy load, and its capability of delivering radio frequency signals to a moving resonator. We employ our system to image vibrations of two FLG resonators. The first resonator is suspended over a nanofabricated local gate electrode acting both as a mirror for optical detection and as a capacitor plate to actuate vibrations at radio frequencies. We measure the hardening of the spring constant of the lowest frequency mode of the resonator as a function of absorbed optical power. From these measurements we estimate the ratio of thermal expansion coefficient to thermal conductivity of the membrane, which plays an important role in thermal transport across the resonator. In addition, imaging the vibrations of this mode allows us to measure its effective mass, which is important for quantitative mass and force sensing applications based on 2-D resonators. Our measurements combine three interesting features which, to the best of our knowledge, have not been reported thus far in a single device: they demonstrate (i) vibration imaging of a lowest frequency mode which (ii) resonates above 60~MHz using (iii) a local metal gate to enhance optical readout. The second resonator has a fold in the membrane that is not visible in an optical microscope. We image the first three vibration modes and show that the fold locally suppresses vibrations. These measurements demonstrate that our system can be used to identify mesoscopic defects in the resonator and study their impact on the mechanical response.

\section{Experimental setup}

\begin{figure}[t]
\centering
\includegraphics{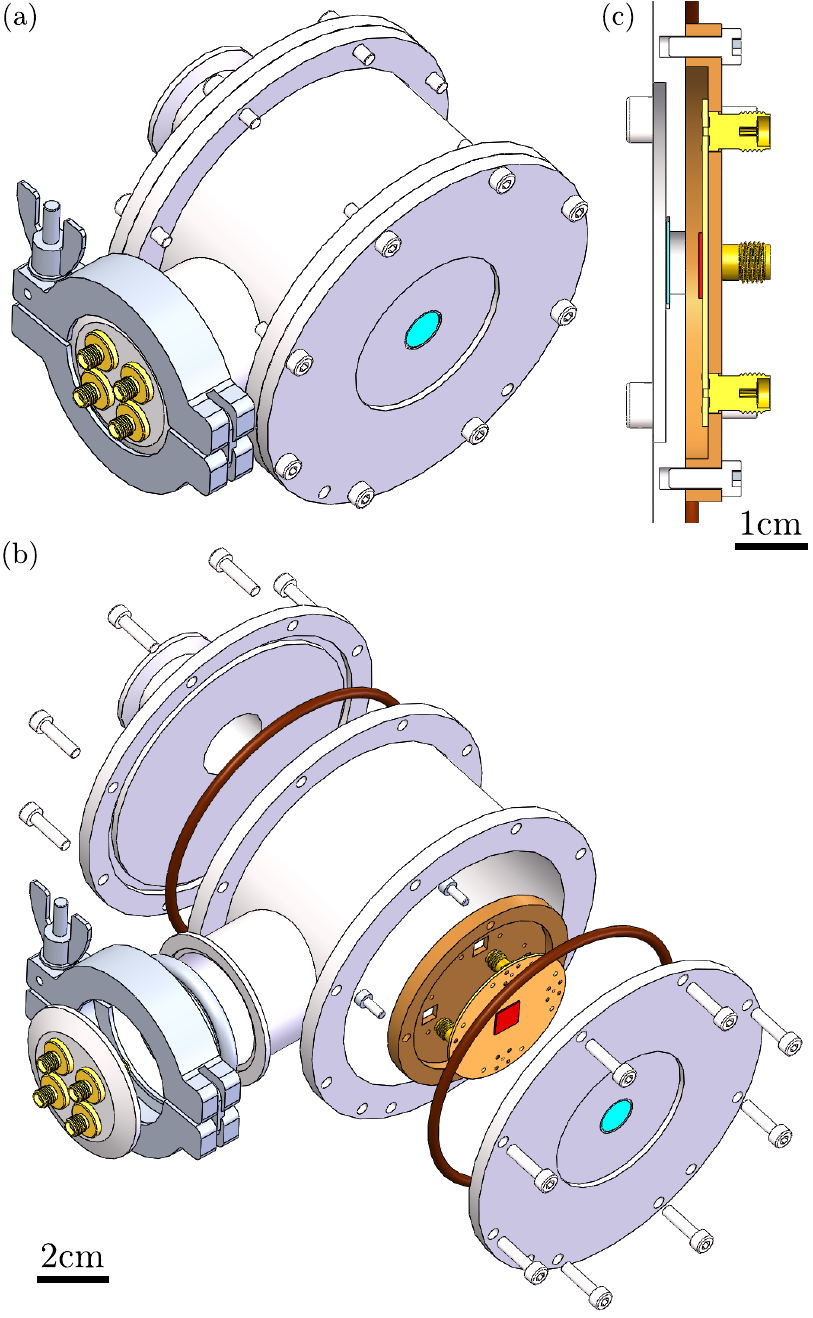}
\caption{Enclosure to image vibrations in vacuum. (a) Assembled enclosure with a window (blue shaded disk) for free space optical measurements, a KF port on the side for SMA connectors and another KF port on the back side for pumping. (b) Exploded view. The red shaded square represents the substrate hosting the resonators. It is glued on a printed circuited board (lightly colored disk) that is attached to a copper holder (brown shaded disk with a shallow recess). (c) Cross section showing the holder (brown), the board (yellow), the substrate (red) and SMA connectors attached to the back side of the board and connected to the front side of it.}\label{enclosure}
\end{figure}

The mechanical part of our system is robust and easy to operate. Our enclosure is shaped as a cylinder with a diameter of 120~mm and a depth of 75~mm, and is made of stainless steel (Fig.~\ref{enclosure}a, b). There is a clear advantage to such a sturdy design: with a total weight of $\approx2.2$~kg, the enclosure acts as an efficient damper for acoustic vibrations which would otherwise preclude certain experiments, such as those where the frequency of the resonator is slowly modulated. To enable free space optical measurements, a window made of fused silica, with a diameter of 12.7~mm and a thickness of 0.4~mm, is glued on the front panel with Torr Seal epoxy. A holder for the substrate hosting the resonators is affixed to the inner side of the front panel facing the window (Fig.~\ref{enclosure}c). The holder accommodates a radio frequency printed circuit board (PCB) that is connected to semi-rigid cables on its back side using SMA connectors soldered through the PCB. The enclosure is sealed with Viton o-rings, enabling a dynamic vacuum as low as $10^{-6}$~mbar using a small size turbomolecular pump connected to a one-meter long KF16 flexible bellow. The enclosure is mounted on a three-axis micromechanical linear stage (Newport M-562-XYZ). Each axis is connected to a manual adjuster (Newport DS-4F) with a specified fine sensitivity of 20~nm. Each adjuster is driven by a stepper motor (Makeblock 42BYG) whose shaft is connected to the adjuster using gears and a rubber belt. In vibration imaging experiments where a laser beam is used to measure the response of a resonator, our system guarantees that only the position of the resonator is varied while the light path remains unchanged, so the incident optical power and the shape of the focused beam are unaffected by the imaging process. Using a nanofabricated calibration sample, we measure the precision of in-plane displacement of the enclosure to be $\approx0.20$~$\mu$m (Supplementary Material, Section I).

\begin{figure}[t]
\centering
\includegraphics{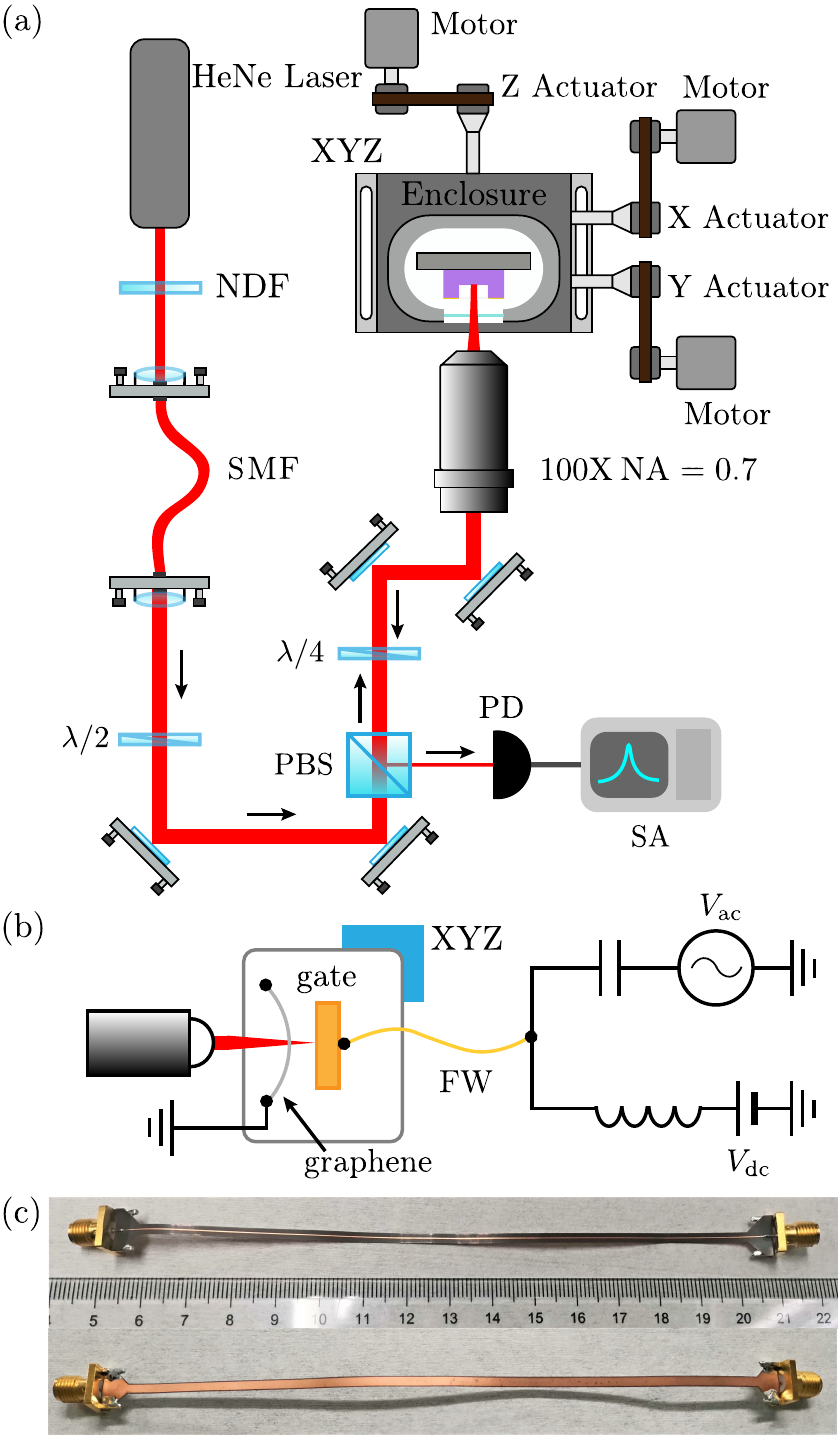}
\caption{Vibration measurement setup. (a) Optical setup. NDF: neutral density filter. SMF: single mode fiber. $\lambda/2$: half-wave plate. PBS: polarizing beam splitter. $\lambda/4$: quarter-wave plate. XYZ: micro-positioner. SA: spectrum analyzer. PD: photodetector. (b) Electrical actuation scheme. FW: flexible waveguide. The gate electrode is connected to a semi-rigid radio-frequency cable inside the enclosure, which is connected to FW outside of the enclosure via a hermetic feed-through. (c) Front and back side of FW consisting of a 15~cm long copper microstrip patterned on Kapton and terminated with SMA connectors.}\label{optical_setup}
\end{figure}

Our optical setup, shown in Fig.~\ref{optical_setup}a, is similar to those used to detect vibrations of 2-D resonators (see \textit{e.g.} Ref.~\cite{graphene1}). Its design originates from the setup presented in Refs.~\cite{Carr,Erkinci}. Briefly, we employ a Helium-Neon laser emitting at a wavelength $\lambda\approx633$~nm as a monochromatic light source. The output of the laser is filtered with a single mode fiber to obtain a clean fundamental transverse Gaussian mode. The combination of a polarizing beam splitter and a quarter-wave plate ensures that light incident on the resonator and reflected light have orthogonal polarizations so the photodetector mostly collects reflected light. Incident light is focused with and reflected light is collected by a long working distance objective (Mitutoyo M Plan Apo 100X) with a numerical aperture $\textrm{NA}=0.7$ ensuring a focused beam size limited by diffraction. Radio frequency voltages are supplied to the resonators (Fig.~\ref{optical_setup}b) via ultra-flexible waveguides (Fig.~\ref{optical_setup}c), as discussed later. We measure the radius $w_0$ of the waist of the focused laser beam using a modified version of the knife edge technique (Supplementary Material, Section II). We find $w_0\approx0.40$~$\mu$m, which is consistent with the input beam parameters.

\section{Results and discussion}

With the beam optimally focused, we demonstrate that our system can image the flexural vibrations of 2-D resonators based on suspended membranes of few layer graphene (FLG). We present vibration imaging data obtained with two devices. The first device is a locally gated FLG resonator. It is characterized by a nanofabricated gate electrode made of evaporated gold over which FLG is suspended. The gate electrode serves as a highly reflective mirror for optical detection, and also forms a capacitor with FLG to actuate vibrations at radio frequencies. We use this device to measure the ratio of thermal expansion coefficient to thermal conductivity of FLG. We also use it to measure the effective mass of the fundamental mode of vibration. The second device is a globally gated FLG resonator. It consists of FLG suspended over a doped silicon substrate. We use this device to demonstrate that our system can be employed to detect the presence of folds in the thin membrane that are otherwise invisible in an optical microscope. The advantage of a local gate made of gold over a global silicon gate is the higher reflectance of the mirror combined with the possibility of actuating individual resonators within an array of devices. To the best of our knowledge, the use of a local gate both for optical detection and capacitive actuation has not been reported thus far.

\begin{figure}[t]
\centering
\includegraphics{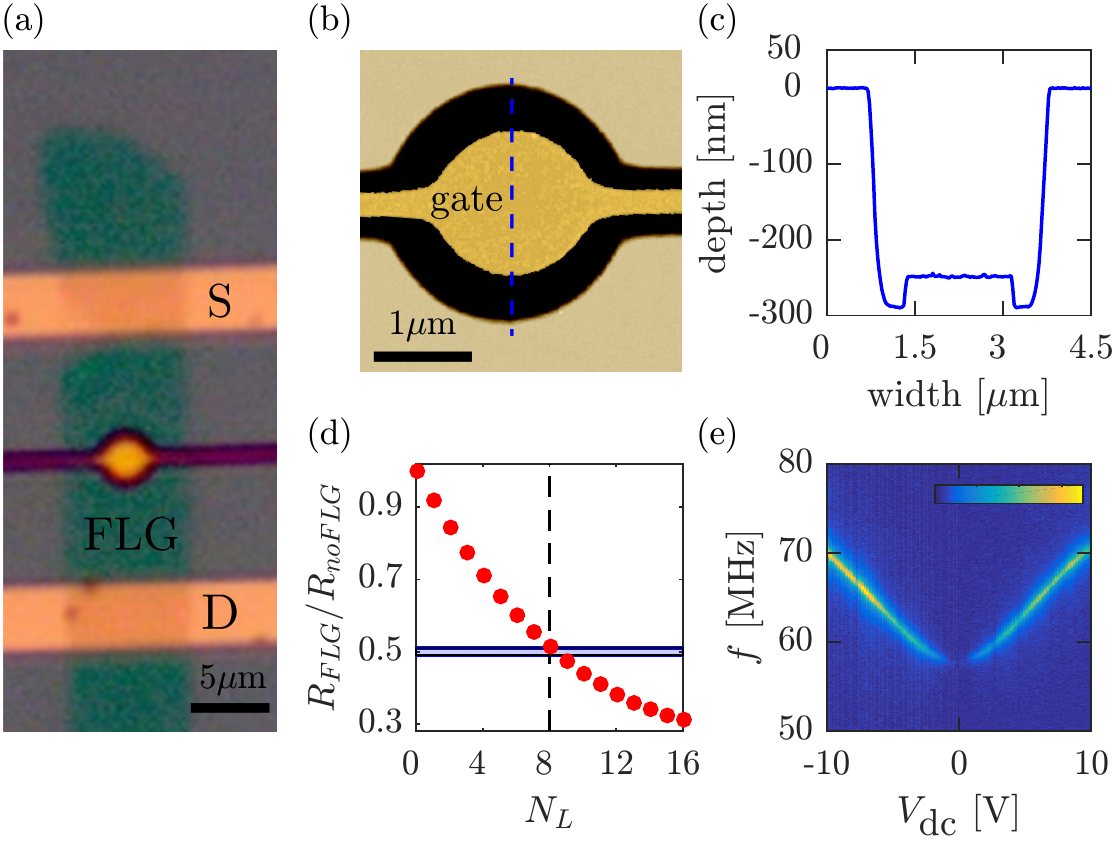}
\caption{Cavity, gate electrode, FLG thickness and tuning of the resonant frequency. (a) Optical microscopy image showing FLG and source (S) and drain (D) electrodes. (b) AFM image of the gate electrode at the bottom of the cavity, and (c) cross section along the blue dashed line. (d) Calculated ratio of reflectances $R_\textit{FLG}/R_\textit{noFLG}$ as a function of the number of layers $N_L$. $R_\textit{FLG}$ is the reflectance of the structure composed of FLG on a 500~nm thick slab of SiO$_2$ on Si at $\lambda=633$~nm, and $R_\textit{noFLG}$ is the reflectance without FLG. The blue shaded area is the uncertainty related to oxide thickness measurements. (e) $\langle V_\textrm{pd}^2\rangle$ as a function of $f$ and $V_\textrm{dc}$ for the lowest frequency mode. $V_\textrm{ac}=0.4$~mV$_\textrm{rms}$, $P_\textrm{inc}=110$~$\mu$W. Orange side of the color bar: $5\times10^{-11}$~V$_\textrm{rms}^2$.}\label{cavity}
\end{figure}

We first consider our locally gated resonator. It is fabricated by exfoliating FLG and transferring it onto a prefabricated substrate using the viscoelastic transfer method \cite{Transfer}. The substrate is thermal silicon oxide grown on highly resistive silicon. It is patterned with source and drain electrodes to contact FLG (Fig.~\ref{cavity}a), a 3~$\mu$m diameter, cylindrical cavity etched in the oxide, and a local gate electrode nanofabricated at the bottom of the cavity (Fig.~\ref{cavity}a, b). The distance between FLG and the gate is nominally 250~nm (Fig.~\ref{cavity}c). We estimate that our FLG is composed of $N_L=8$ graphene layers from measurements of optical power reflected by the oxidized silicon substrate with and without supported FLG, away from the cavity. We obtain $N_L$ by comparing the ratio of these measured powers to calculations based on the transfer matrix method \cite{Roddaro,FengNan}, see Fig.~\ref{cavity}d and Supplementary Material, Section~III. Additional reflected power measurements made on the gold electrodes confirm this result. The resonator is actuated electrically by applying an oscillating voltage $V_\textrm{ac}$ superimposed on a dc offset $V_\textrm{dc}$ between FLG and the gate (Fig.~\ref{optical_setup}b). This results in an electrostatic force of amplitude $V_\textrm{dc}V_\textrm{ac}|\frac{\mathrm{d}C}{\mathrm{d}z}|$, where the third factor is the derivative with respect to flexural displacement of the capacitance $C$ between FLG and the gate. We favor electrostatic drive over optical drive \cite{graphene1} because the former allows actuating very high frequency vibrations without the inconvenience of dissipation caused by photothermal effects. To supply radio frequency signals for this actuation without impeding the motion of the micropositioning stage, we use a homemade waveguide consisting of a copper microstrip patterned on a 15~cm long ribbon cut from a thin Kapton film (Fig.~\ref{optical_setup}c). Our waveguide is ultra-flexible, its insertion loss is smaller than 1~dB up to at least 3~GHz, and its scattering parameters are insensitive to bending and twisting of the waveguide. The resonator is kept at room temperature in a vacuum of $\approx10^{-6}$~mbar. We detect the flexural vibrations of the resonator using a standard technique \cite{graphene1,Carr,Erkinci}. Briefly, the resonator is placed in an optical standing wave from which it absorbs energy. Vibrations render absorbed energy time-dependent, resulting in modulations of the reflected power. Correspondingly, the mean square voltage $\langle V_\textrm{pd}^2\rangle$ at the output of the photodetector reads
\begin{equation}
\langle V_\textrm{pd}^2\rangle\approx\left(G\times T\times P_\textrm{inc}\Big|\frac{\mathrm{d}R}{\mathrm{d}z}\Big|_{z_M}\right)^2\langle z_\textrm{vib}^2\rangle+\langle\delta V_\textrm{b}^2\rangle\,,\label{Pm}
\end{equation}
where $G$, in units of V/W, is the product of the responsivity and the transimpedance gain of the photodetector, $T$ is the transmittance of the reflected light path, $P_\textrm{inc}$ is the optical power incident on the resonator, $z_\textrm{vib}$ is the amplitude of vibrations in the flexural direction, $\delta V_\textrm{b}$ is the amplitude of fluctuations of the measurement background, and $\langle\cdot\rangle$ averages over time. The quantity $|\frac{\mathrm{d}R}{\mathrm{d}z}|_{z_M}$ is the derivative of the reflectance $R$ of the whole device consisting of the resonator and the reflective gate, at a distance $z_M$ between the resonator and the gate. The local gate acts as highly reflective mirror that optimizes the transduction of $\langle z_\textrm{vib}^2\rangle$ into $\langle V_\textrm{pd}^2\rangle$. We calculate $|\frac{\mathrm{d}R}{\mathrm{d}z}|\approx5\times10^{-3}$/nm for our device at $\lambda=633$~nm, which is about twice the value calculated with a regular silicon gate \cite{Roddaro}. To study the frequency response $\langle z_\textrm{vib}^2\rangle(f)$ of the resonator, we sweep the drive frequency $f$ of $V_\textrm{ac}$ and measure $\langle V_\textrm{pd}^2\rangle$ with a spectrum analyzer. Figure~\ref{cavity}e displays $\langle V_\textrm{pd}^2\rangle$ as a function of $f$ and $V_\textrm{dc}$ for the lowest frequency mode we can resolve. The response shifts to higher frequencies as $|V_\textrm{dc}|$ increases mostly because the electrostatic force $\propto|\frac{\mathrm{d}C}{\mathrm{d}z}|V_\textrm{dc}^2$ tensions the membrane as it pulls it towards the gate, making the resonant frequency of the mode tunable.

\begin{figure}[t]
\centering
\includegraphics{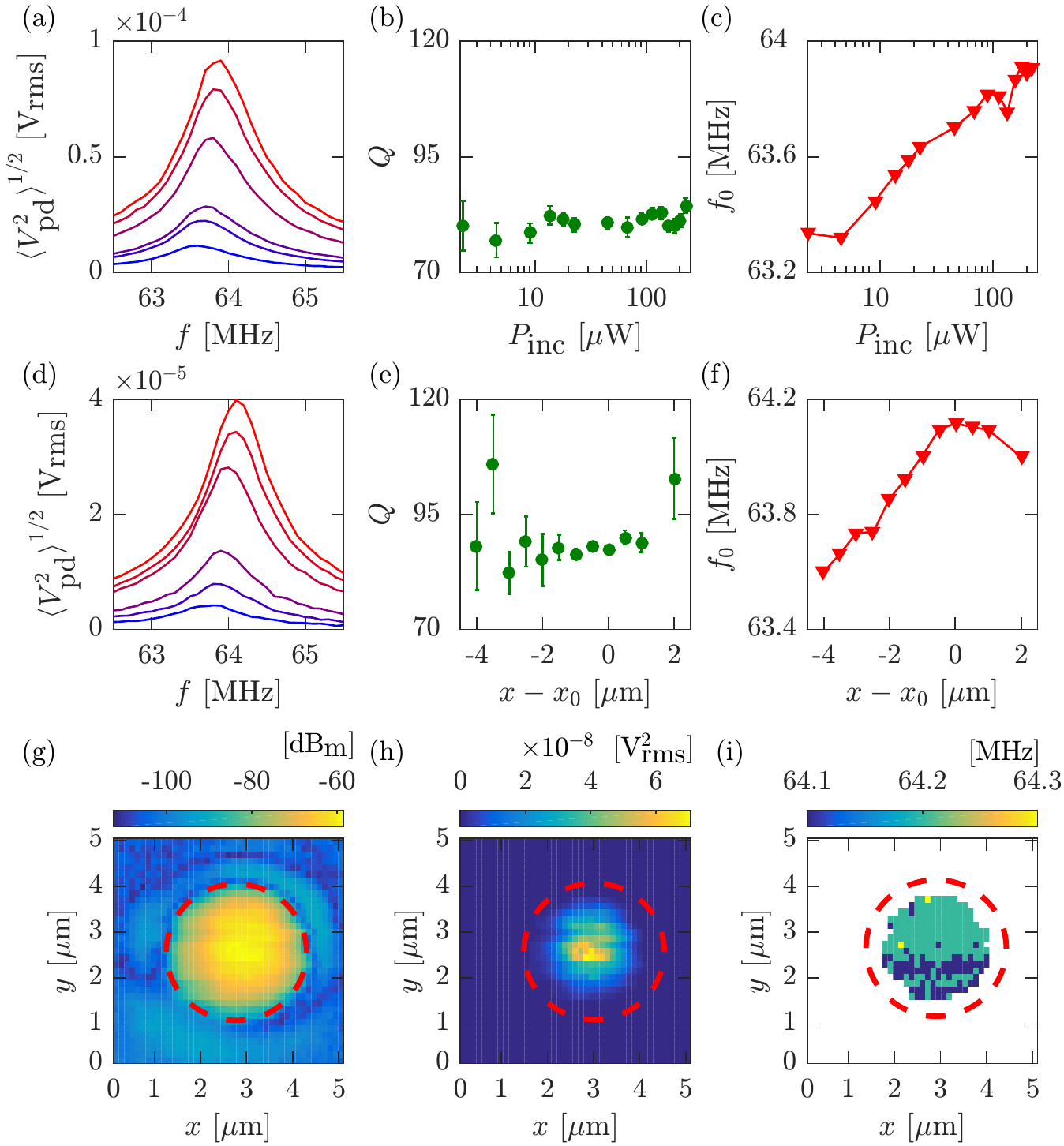}
\caption{Effect of the beam on the response of the resonator and vibration imaging. (a) Root mean square of the voltage at the output of the photodetector, $\langle V_\textrm{pd}^2\rangle^{1/2}$ as a function of drive frequency $f$ at $V_\textrm{ac}=12.6$~mV$_\textrm{rms}$ and $V_\textrm{dc}=5$~V with the resonator near the center of the beam. From blue to red: $P_\textrm{inc}=20$, $45$, $70$, $135$, $160$, and $205$~$\mu$W. (b) Quality factors $Q$ and (c) resonant frequency $f_0$, measured from a series of resonances partially shown in (a), as a function of $P_\textrm{inc}$. (d) $\langle V_\textrm{pd}^2\rangle^{1/2}$ as a function of $f$ as the resonator is moved along $x$ into the beam at $x_0$ ($V_\textrm{ac}=12.6$~mV$_\textrm{rms}$, $V_\textrm{dc}=5$~V and $P_\textrm{inc}=110$~$\mu$W). (e) $Q$ and (f) $f_0$ as a function of in-plane displacement $x-x_0$ taken from resonances partially shown in (d). (g) Peak (resonant) value of the time averaged electrical power $\langle V_\textrm{pd}^2\rangle/50$ as a function of $x$ and $y$. (h) Peak value of the electromechanical signal $\bar{V}^2$, shown here on a linear scale. $V_\textrm{dc}=5$~V, $V_\textrm{ac}=0.4$~mV$_\textrm{rms}$ and $P_\textrm{inc}=110$~$\mu$W. (i) Peak frequency of $\bar{V}^2(f)$. Data in (g)-(i) were obtained after optimizing the collection efficiency of the photodetector compared to data in (a)-(f).}\label{figXY}
\end{figure}

With the resonator positioned near the center of the focused beam, we characterize the lowest frequency resonance in the spectrum of $\langle V_\textrm{pd}^2\rangle(f)$ and its dependence on $P_\textrm{inc}$. We have verified that, on resonance, the electromechanical signal represented by the root mean square voltage $\bar{V}=(\langle V_\textrm{pd}^2\rangle-\langle\delta V_\textrm{b}^2\rangle)^{1/2}$ increases linearly with $V_\textrm{ac}$ within the range used in this work while the resonant frequency does not shift, indicating that the resonator is driven in a regime where the restoring force is linear in displacement.  Figure~\ref{figXY}a shows $\langle V_\textrm{pd}^2\rangle^{1/2}(f)$ measured at $V_\textrm{ac}=12.6$~mV$_\textrm{rms}$ and $V_\textrm{dc}=5$~V for $P_\textrm{inc}$ ranging from 20~$\mu$W to 205~$\mu$W. Here as well, we have verified that the peak value of $\bar{V}(f)$ increases linearly with $P_\textrm{inc}$. Overall, $\bar{V}$ can be linearly amplified either by increasing the vibrational amplitude electrostatically with $V_\textrm{ac}$ or by increasing the optical readout with a larger probe power $P_\textrm{inc}$. The lineshape of $\bar{V}^2(f)$ is Lorentzian, which indicates that the resonator behaves as a damped harmonic oscillator with susceptibility
\begin{equation}
\chi(f)=\frac{1}{4\pi^2}\frac{1}{f_0^2-f^2-\mathrm{i}f_0f/Q}\,,\label{response}
\end{equation}
where $Q$ is the spectral quality factor, $f_0$ is the resonant frequency of the vibrational mode, and $\bar{V}^2(f)\propto|\chi(f)|^2$. Figure~\ref{figXY}b shows that $Q$ is low and does not change within the range of $P_\textrm{inc}$, while $f_0$ increases with $P_\textrm{inc}$ (Fig.~\ref{figXY}c). In the absence of nondissipative spectral broadening processes, $Q$ is inversely proportional to the rate at which energy stored in a resonator gets dissipated in a thermal bath. In low dimensional resonators based on 2-D materials and on nanotubes, $Q$ at room temperature is always found to lie between 10 and $\approx100$ \cite{graphene1,Sazanova}, which is surprisingly low given the high crystallinity of the resonators. Proposed mechanisms to explain such low $Q$'s include losses within the clamping area \cite{clamping} and spectral broadening due to nonlinear coupling between the mode of interest and a large number of thermally activated modes \cite{Barnard,Yaxing2015}. In turn, the increase of $f_0$ reveals a hardening of the spring constant of the resonator. The latter may be due to vibrations responding to photothermal forces with a delay \cite{graphene3,Karrai,Favero,Buks}. However, because $Q$ does not appreciably change with $P_\textrm{inc}$, the hardening of the spring constant is more likely to be caused by absorptive heating accompanied by a contraction of the membrane \cite{Davidovikj,Yoon}. In this case, it is interesting to relate the change $\Delta f_0$ induced by a change $\Delta P_\textrm{inc}$ to the thermal expansion coefficient $\alpha$ and to the thermal conductivity $\kappa$ of the membrane. For this we borrow a result from Ref.~\cite{Nicolas}, namely $|\Delta f_0/\Delta P_\textrm{abs}|=|\alpha f_0\eta/(4\pi\epsilon\kappa h)|$, with $P_\textrm{abs}$ the power absorbed by the membrane, $\epsilon$ the strain within the membrane, $h=N_L\times0.34\times10^{-9}$~m the thickness of the membrane, and $\eta\approx1$ a factor that depends on the beam radius, on the membrane radius and on Poisson's ratio. We convert $P_\textrm{inc}$ into $P_\textrm{abs}$ using the absorbance $A=P_\textrm{abs}/P_\textrm{inc}$ of our FLG suspended over the gate electrode. We measure $A\approx0.3$ from the ratio of power reflected by the cavity covered by FLG to the power reflected by a nearby uncovered cavity (Supplementary Material, Section~III). Further, we estimate $\epsilon\approx5\times10^{-4}$ from $f_0$ by calculating the elastic energy of a disk-shaped membrane \cite{Timoshenko,Yin} and deriving from it the spring constant of the fundamental mode (Supplementary Material, Section~IV). We make the simplifying assumption that the electrostatic force is uniform over the membrane. We also assume a Young's modulus of $10^{12}$~Pa and a Poisson's ratio of 0.165 \cite{Blakslee}. Combining $\Delta f_0/\Delta P_\textrm{inc}$, $A$ and $\epsilon$, we find $|\alpha/\kappa|\approx4\times10^{-9}$~m/W. This is a reasonable estimate, considering for example $|\alpha/\kappa|\approx4\times10^{-9}$~m/W with $\alpha\approx-8\times10^{-6}$~K$^{-1}$ from suspended singe layer graphene \cite{Yoon} and $\kappa\approx2000$~Wm$^{-1}$K$^{-1}$ from pyrolytic graphite \cite{Pyrolytic}, both at room temperature.

We now measure the response of the resonator as a function of its in-plane position with respect to the beam. Data shown in Figs.~\ref{figXY}d-f are measured with $V_\textrm{ac}=12.6$~mV$_\textrm{rms}$, $V_\textrm{dc}=5$~V and $P_\textrm{inc}=110$~$\mu$W. Figure~\ref{figXY}d shows the same resonance as in Fig.~\ref{figXY}a measured at various fixed positions along the $x$ direction, with the center of the beam at $x_0$. Correspondingly, Fig.~\ref{figXY}e shows $Q$ and Fig.~\ref{figXY}f shows $f_0$ as a function of $x-x_0$. Figures~\ref{figXY}d-f are strikingly similar to Figs.~\ref{figXY}a-c: $Q$ does not depend on position while $f_0$ increases as the resonator approaches $x_0$ and decreases as it moves out of the beam. As the resonator moves with respect to the position of the beam, it samples the intensity of the beam in a similar way to that of a reflective structure in a knife edge measurement (Supplementary Material, Section II). Importantly, the dependence of $f_0$ on position means that vibration imaging requires measuring the full resonance at each position on the resonator, as we do next.

We present spatially resolved amplitude measurements of the lowest frequency mode in Figs.~\ref{figXY}g-i. Figure~\ref{figXY}g shows the peak (resonant) value of the time averaged electrical power $\langle V_\textrm{pd}^2\rangle/50$ dissipated across the input impedance of the spectrum analyzer, expressed in dB$_\textrm{m}$, as a function of $x$ and $y$. Figure~\ref{figXY}h shows the peak value of $\bar{V}^2$ on a linear scale. The latter is proportional to the mean square of the resonant vibrational amplitude, see Eq.~(\ref{Pm}), hence it is proportional to the potential energy of the mode. The background $\langle\delta V^2_\textrm{b}\rangle$ is measured in the cavity at $f_0$ but with the drive frequency shifted up and far away from resonance. Measurements are made at $V_\textrm{dc}=5$~V with $V_\textrm{ac}=0.4$~mV$_\textrm{rms}$ and $P_\textrm{inc}=110$~$\mu$W. At such low $V_\textrm{ac}$, we find that the peak value of $\bar{V}^2$ is sizeable only in a central area away from the edge of the cavity (highlighted by the dashed circle in Fig.~\ref{figXY}h). Within this area, the measured peak frequency of $\bar{V}^2(f)$ is almost uniform and defines the resonant frequency $f_0$ of the mode (Fig.~\ref{figXY}i).

We use the vibration imaging data shown in Figs.~\ref{figXY}g-i to measure the effective mass $m_\textrm{eff}$ of the mode. The latter is related to the potential energy $U$ as
\begin{equation}
U\equiv\frac{1}{2}\frac{m}{\pi a^2}(2\pi f_0)^2\iint z_\textrm{vib}^2(f_0,x,y)\mathrm{d}S=\frac{1}{2}m_\textrm{eff}(2\pi f_0)^2z_\textrm{max}^2\,,\label{meff}
\end{equation}
where $m$ is the geometrical mass and $a$ is the radius of the membrane, respectively, $z_\textrm{vib}(f_0,x,y)$ is the resonant amplitude at position $(x,y)$, $\mathrm{d}S$ is an elementary area on the membrane and $z_\textrm{max}$ is the largest value of $z_\textrm{vib}(f_0,x,y)$ over the membrane. Replacing the integral in Eq.~(\ref{meff}) with a discrete sum yields
\begin{equation}
\frac{m_\textrm{eff}}{m}=\frac{\sum_{i,j}\bar{V}^2(f_0,x_i,y_j)}{N\bar{V}^2_\textrm{max}}\,,\label{meffdiscrete}
\end{equation}
where $x_i$ and $y_j$ are discrete coordinates over the cavity, $N$ is the number of pixels within the area of the cavity in Figs.~\ref{figXY}g-i, and $\bar{V}^2_\textrm{max}$ is the largest value of $\bar{V}^2(f_0,x_i,y_j)$ over the membrane. While Figs.~\ref{figXY}g-i represent the convolution of the focused beam with the vibration mode shape (instead of the mode shape alone), we verified numerically that $m_\textrm{eff}$ calculated from the convolution and $m_\textrm{eff}$ calculated from the mode shape agree within 5\% for our measured radius $w_0\approx0.4$~$\mu$m of the waist of the focused laser beam. Equation~(\ref{meffdiscrete}) yields $m_\textrm{eff}/m=0.27\pm0.01$. This estimate agrees well with the value of 0.27 calculated for a disk shaped graphene membrane without bending rigidity and subjected to electrostatic pressure \cite{Peter}. It shows that the assumption of negligible bending rigidity compared to stretching rigidity is still a valid one for FLG.

We complement our study with vibration imaging measurements performed on a second device, showing the effect of a mesoscopic defect in FLG on the mechanical response. Here the resonator consists of FLG suspended over silicon oxide grown on doped silicon. The silicon substrate serves as a global gate electrode. We show the mechanical resonance spectrum of the resonator and its dependence on $V_\textrm{dc}$ in Supplementary Material, Section~V. Figure~\ref{figSEM}a is a scanning electron microscope image of the device, which reveals the presence of a fold in FLG near the bottom edge of the cavity. This fold cannot be seen in an optical microscope with a $100\times$ magnification objective. Figures~\ref{figSEM}b-d display the resonant value of $\langle V_\textrm{pd}^2\rangle$ as a function of in-plane displacements $x$ and $y$ for the first three vibrational modes we are able to measure. While these results are qualitatively consistent with the shapes of a first, second and third mode, we observe the presence of a node in the vicinity of the fold. The fold presumably causes a local stiffening of FLG \cite{graphenedot} which pins vibration modes and forces them to a low amplitude state. Our measurements show that folds have a strong impact on the mechanical response of 2-D resonators. As with membranes with free standing edges \cite{embracing} and membranes with inhomogeneous strain and mass distributions \cite{Daniel}, membranes with folds may have mechanical properties that may not be found in uniform membranes. Our measurement system is well suited to investigate these properties as it is noninvasive and, unlike scanning electron imaging, does not contaminate the surface of resonators.

\section{Conclusion}
Our simple system composed of a vacuum enclosure mounted on a three-axis micropositioning stage is well suited to measure the mechanical response of few layer graphene electromechanical resonators and to image their vibrations. From the hardening of the spring constant of the lowest frequency mode in response to increased incident power, we estimate the ratio of thermal expansion coefficient $\alpha$ to thermal conductivity $\kappa$ of the membrane. Doing so requires either a strong temperature gradient induced by the beam across the membrane or a resonator with a low spring constant, both of which are more likely to be obtained with resonators larger than our 3~$\mu$m diameter device \cite{graphene3,Davidovikj}. We image the shape of the lowest frequency mode and measure its effective mass, which is important for quantitative sensing applications based on those devices. We also image vibration modes in the presence of a fold in FLG, and show that the fold strongly affects the mechanical response of the resonator. Built-in calibration of in-plane displacement is possible by mapping the reflectance of the cavity, which can be done by averaging the measurement background away from resonance in between two resonant measurements. Our system may be used in combination with a small size, dry cryostat that would replace our heavy vacuum enclosure and with the objective outside the cryostat. Measuring the dependence of $f_0$ on temperature would yield $\alpha$ which, combined with $f_0(P_\textrm{inc})$, would yield $\kappa$ \cite{Nicolas}. Planning for such experiments, we have experimentally verified that our homemade waveguides remain flexible and that their insertion loss remains low at cryogenic temperatures by bending 4 of them with a piezopositioner at 800~mK. If precision and accuracy on the nanometer scale are not needed, our system may offer an alternative to systems based on piezo positioners which are fragile, have a small load capacity, and often come at a prohibitive cost to experimentalists on a budget.

\section*{Acknowledgments}
J. Moser is grateful to Yin Zhang, Warner J. Venstra and Alexander Eichler for helpful discussions. This work was supported by the National Natural Science Foundation of China (grant numbers 61674112 and 62074107), the International Cooperation and Exchange of the National Natural Science Foundation of China NSFC-STINT (grant number 61811530020), Key Projects of Natural Science Research in JiangSu Universities (grant number 16KJA140001), the project of the Priority Academic Program Development (PAPD) of Jiangsu Higher Education Institutions, and the Opening Fund of State Key Laboratory of Nonlinear Mechanics in Beijing.
\bibliography{Heng_precision2}
\newpage

\newpage

\newpage

\newpage

\newpage
\begin{figure}[t]
\centering
\includegraphics{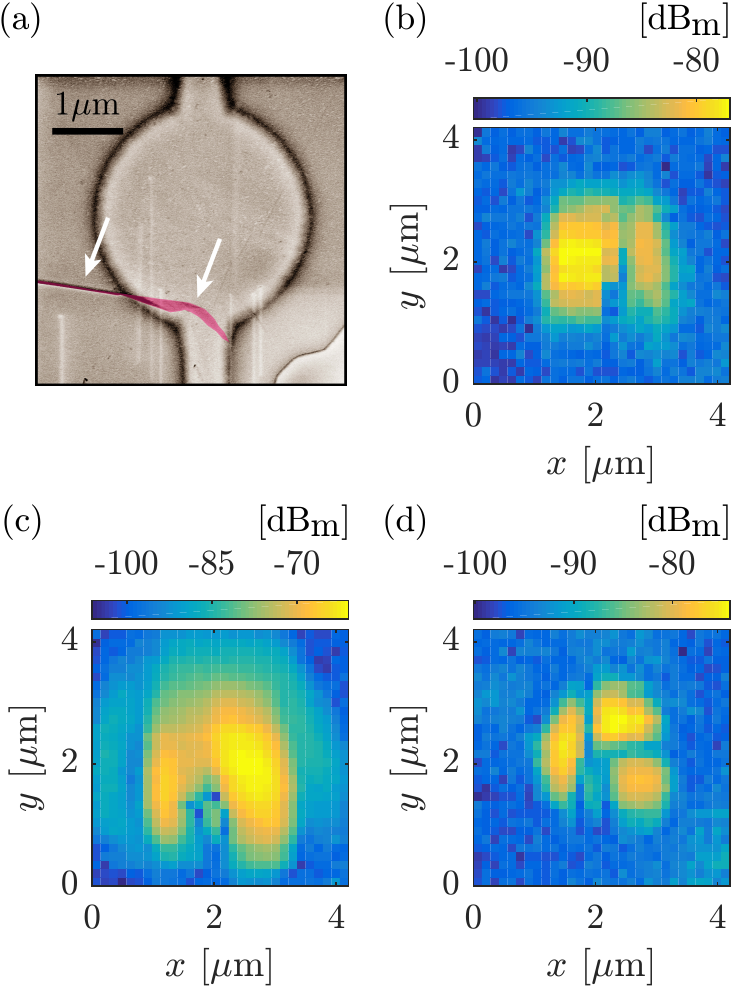}
\caption{Effect of a fold on a the mode shapes of the globally gated resonator. (a) Scanning electron microscope image of the device. The arrows point to a fold in FLG (red shading) near the bottom edge of the cavity. (b-d) $\langle V_\textrm{pd}^2\rangle/50$ as a function of in-plane displacements $x$ and $y$ for the lowest frequency mode near 75~MHz (a), for the second mode near 100~MHz (b) and for the third mode near 150~MHz (d). $V_\textrm{ac}=7.07$~mV$_\textrm{rms}$ is used in (b) and (c) and $V_\textrm{ac}=22.36$~mV$_\textrm{rms}$ is used in (d). $V_\mathrm{dc}=5$~V and $P_\textrm{inc}=110$~$\mu$W are used in all panels.}\label{figSEM}
\end{figure}

\end{document}